\documentclass[twoside]{LCWS11}
\usepackage[latin1]{inputenc}
\usepackage{wrapfig,rotating}
\usepackage{amssymb,amsmath,array}
\usepackage{cite}

\usepackage{amssymb}
\usepackage{amsmath}
\usepackage{units}

\begin{document}
\title{
Beam Test with a GridGEM TPC Prototype Module} 

\author{Ralf Diener$^1$, Ties Behnke$^1$,  Stefano Caiazza$^1$\thanks{Marie Curie Initial Training Network on Particle Detectors (MC-PAD) - PITN-2008 - 214560.},  Isa Heinze$^1$,  Volker Prahl$^1$,  Christoph Rosemann$^1$,\\  Oliver Sch{\"a}fer$^{1,2}$, Jan Timmermans$^{1,3}$, Robert Volkenborn$^1$ and Klaus Zenker$^1$
\vspace{.3cm}\\
1- Deutsches Elektronen-Synchrotron DESY, A Research Centre of the Helmholtz Association\\
Notkestrasse 85, 22607 Hamburg - Germany
\vspace{.1cm}\\
2- Universit{\"a}t Rostock, Fachbereich Physik,  \\
Universit{\"a}tsplatz 3, 18051 Rostock - Germany
\vspace{.1cm}\\
3- Nikhef, National Institute for Subatomic Physics\\
P.O. Box 41882, 1009 DB Amsterdam - Netherlands\\
}

\maketitle

\begin{abstract}
The International Large Detector (ILD) ---a detector concept for the International Linear Collider (ILC)--- foresees a Time Projection Chamber (TPC) as its main tracking detector. Currently, the R\&D efforts for such a TPC focus on studies using a large prototype that can accommodate up to seven read-out modules which are comparable to the ones that would be used in the final ILD TPC. The DESY TPC group has developed such a module using GEMs for the gas amplification, which are mounted on thin ceramic frames. The module design and first results of a test beam campaign are presented.

\end{abstract}

\section{Introduction}

The ILD concept\cite{:2010zzd} foresees a Time Projection Chamber (TPC) as the main tracking detector. This TPC will have a very accurate momentum resolution of $\unit[9 \times 10^{-5}]{/GeV/c}$ at the planned magnetic field of \unit[3.5]{T} and ensures with up to over 200 space points per track a robust tracking. In addition, the TPC is capable of measurements of the specific energy loss which serves as input for particle identification. The material budget is planned to stay below \unit[5]{\%} of a radiation length $X_0$ in the barrel region and below \unit[25]{\%} of $X_0$ at the end caps~\cite{ref-tpcprc2010}. The end plates of the TPC will be equipped with read-out modules using Micro Pattern Gaseous Detectors (MPGD) for gas amplification.

\section{Test beam measurement setup}

\begin{wrapfigure}{r}{0.40\columnwidth}
 \centerline{\includegraphics[width=0.31\columnwidth]{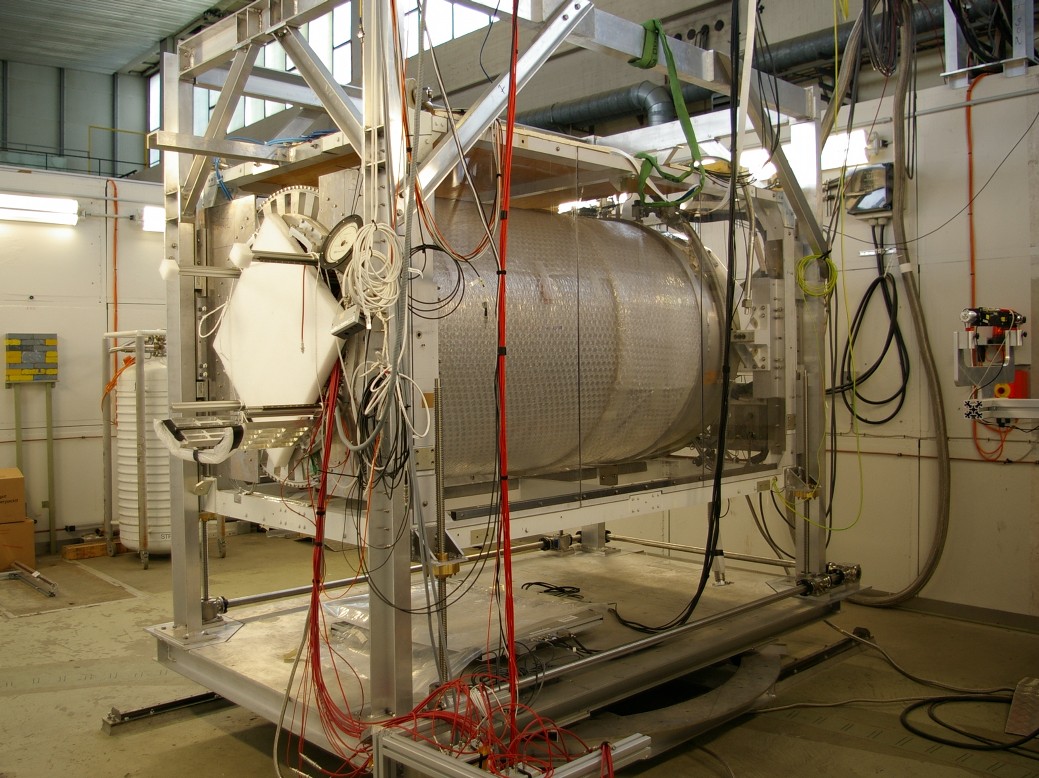}}
\caption [T24 test beam area]{T24/1 test beam area.}
\label{fig:t24}
\end{wrapfigure}

The LCTPC\cite{ref-lctpc} collaboration formed to pursue the research and development of a high-performance TPC for the ILD. Inside the EUDET~\cite{ref-eudet} project, the collaboration has built up a test stand at the DESY test beam, where electron and positron beams from 1 to \unit[6]{GeV} are available (Figure \ref{fig:t24}). The area hosts a sophisticated infrastructure to operate a gaseous detector. The setup comprises the 1T magnet PCMAG mounted on a movable stage and a high voltage and a gas system including a slow control setup. A photo electron calibration system is also available. For the future, the inclusion of silicon detector layers from the former Zeus vertex detector is planned to add the capability to perform reference measurements of the beam position.

The Large Prototype (LP) TPC~\cite{Behnke:2010ze} has been designed to fit into the PCMAG. One of its goals is to study the technical challenges, such as how to scale the read-out structures developed in tests with small prototypes. This includes designs that allow a large area coverage with a minimum amount of dead space.

The anode side end plate of the Large Prototype has been designed by the University of Cornell~\cite{ref-endplate} to house seven read-out modules of a shape and size comparable to one that could be used in the ILD TPC. They are arranged in three rows in the end plate with a curvature that corresponds to the radius of the outermost ring in the ILD TPC.

\section{GridGEM read-out module}

Traditionally, external frames are used to mount GEMs. This technique is not optimal for a large area read-out system. On the one hand, the frame introduces a quite large insensitive area. On the other hand, it is difficult to ensure a good GEM foil flatness by stretching in such a frame. Therefore, a novel mounting technique has been developed at DESY, in which the GEMs are glued to thin ceramic grids made of aluminum oxide. This mounting minimizes the insensitive area with almost edgeless module borders and allows the construction of self supporting integrated GEM read-out modules. Further, the system has a reduced material budget compared to the traditional GRP framing and a flat GEM mounting can be achieved easily. The improved flatness leads to less gain variations and an improved field homogeneity in the drift volume close to the read-out plane. 

\begin{figure}[hbt!]
\centering
\includegraphics[height=0.26\textwidth]{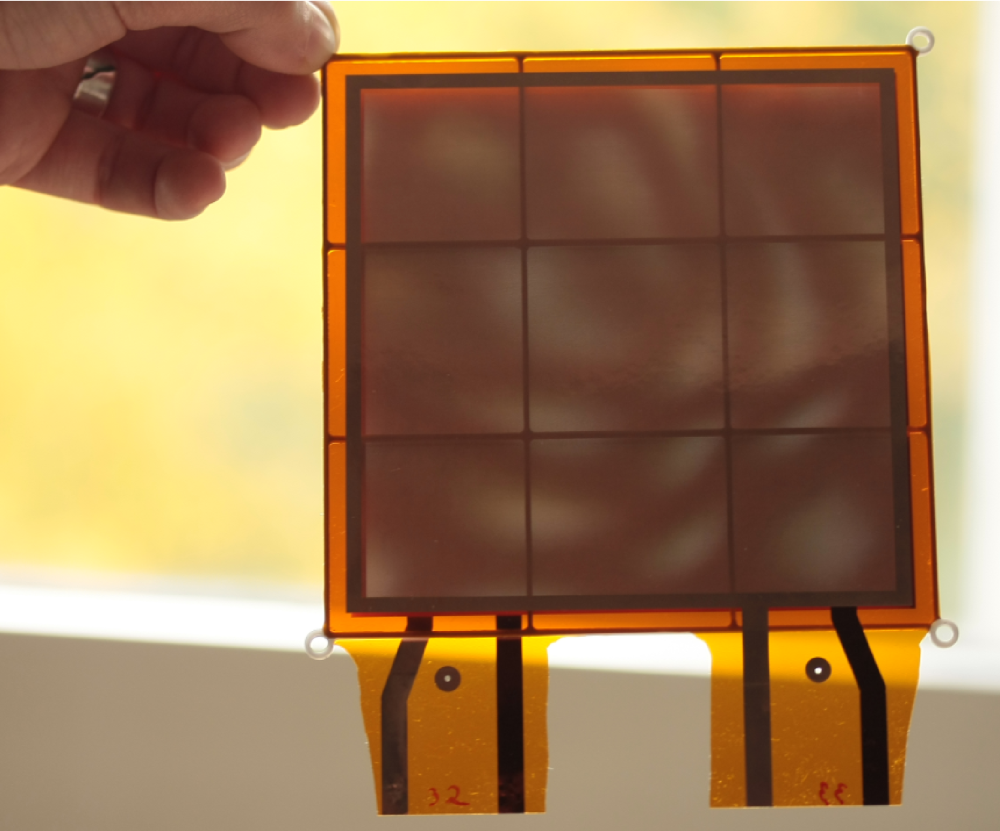}
\hspace{0.05\textwidth}
\includegraphics[height=0.26\textwidth]{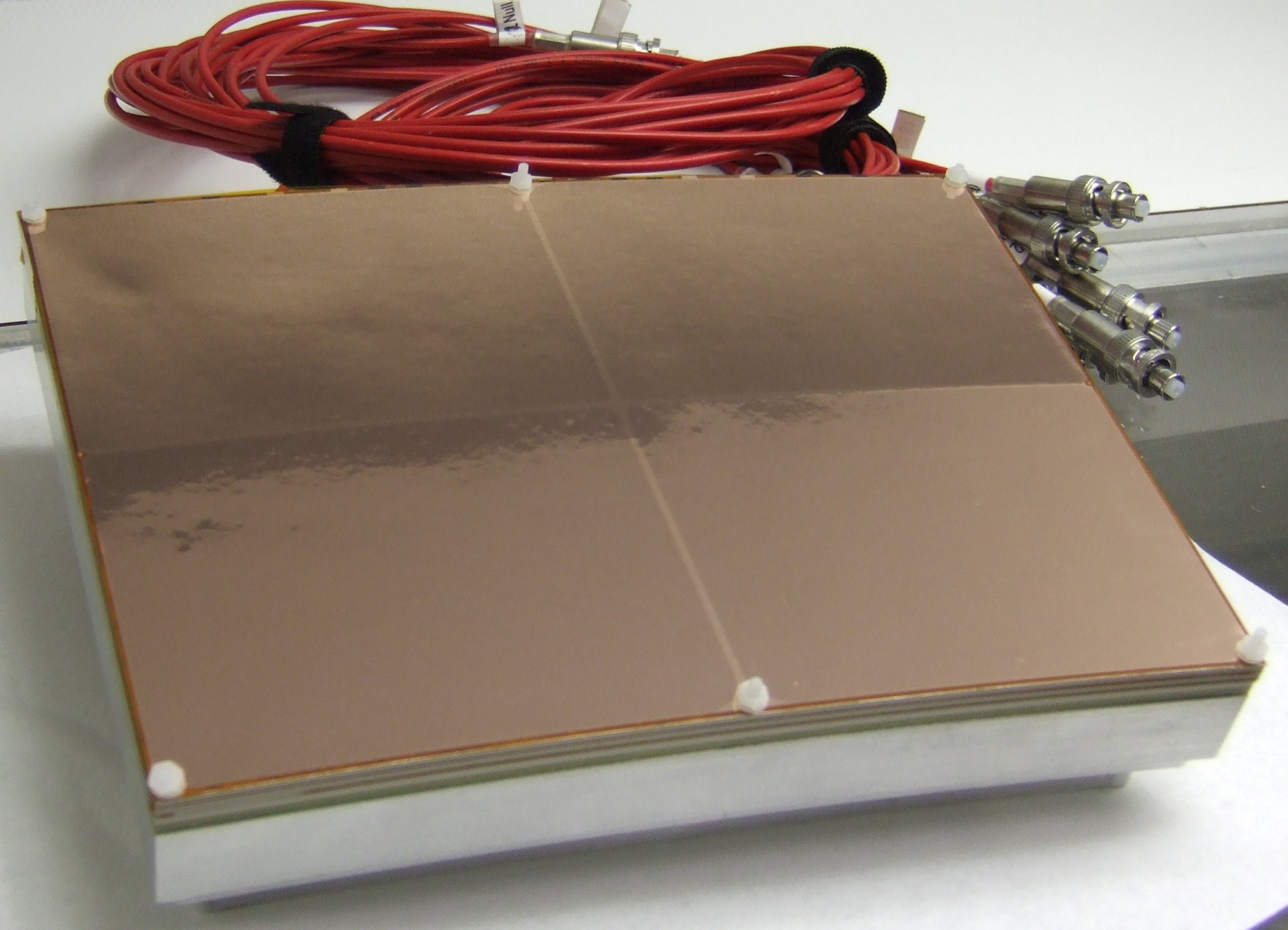}
\caption [GridGEMs]{Left: Standard GEM mounted on a ceramic grid. Right: GridGEM module.}
\label{fig:gridgems}
\end{figure}

After previous studies~\cite{Hallermann:2010zz} with \unit[$10\times10$]{cm$^2$} standard CERN GEMs (left picture of Figure \ref{fig:gridgems}) in a small prototype, the ceramic grid mounting has been implemented in a Large Prototype read-out module (right picture of Figure \ref{fig:gridgems}). The module has a size of about \unit[$23\times17$]{cm$^2$} and uses a triple GEM amplification structure. The ceramic grid divides the area in four sectors. These sectors are electrically separated on one side of the GEMs to limit the stored charge to avoid damaging the GEMs in case of a discharge. The other side of the GEMs has only one sector covering the whole area.

\begin{figure}[hbt!]
\centering
\includegraphics[height=0.25\textwidth]{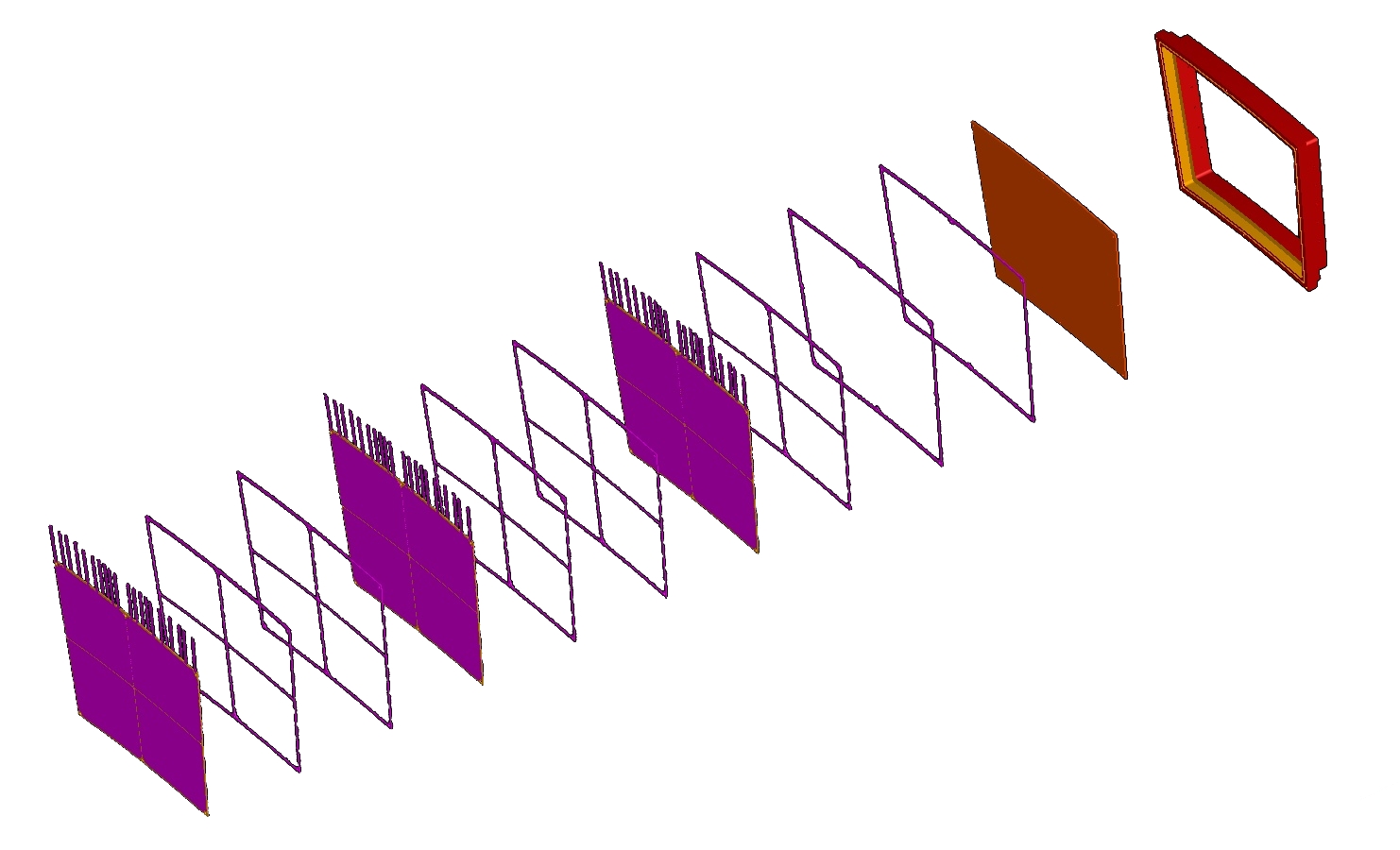}
\hspace{0.05\textwidth}
\includegraphics[height=0.25\textwidth]{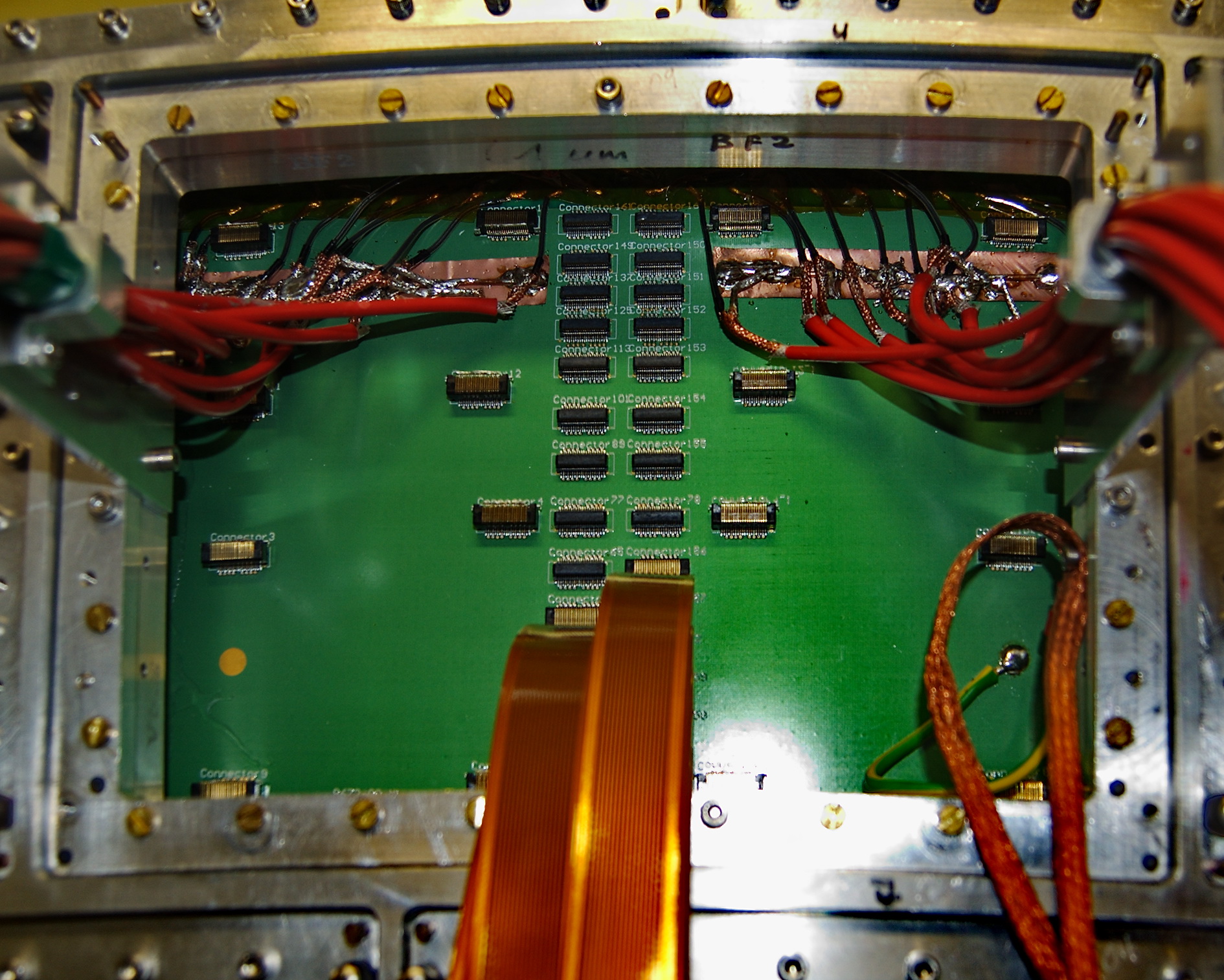}
\caption [Module]{GridGEM module. Left: Exploded assembly drawing. Right: Back side.}
\label{fig:gridgemmodule}
\end{figure}

Figure \ref{fig:gridgemmodule} shows an exploded assembly drawing of the module and a view of the back of the module showing the high voltage and read-out connections. The pad plane was designed by the TPC group at the University of Bonn within the Helmholtz Alliance ``Physics at the Terascale''~\cite{ref-terascale}. In this iteration of the module, the pad plane has a rather simple design in which only the central area was equipped with small read-out pads with a pitch of \unit[$1.26\times5.85$]{mm$^2$}. The rest of the area is equipped with eight times larger, grounded pads allowing for a simpler routing in the board. 

\section{Test beam measurement}

In June and July 2011, the GridGEM module was tested for two weeks in the Large Prototype at the DESY test beam. The module was read out using a modified
ALTRO~\cite{altro} electronics with a PCA16 preamplifier~\cite{pca16} at \unit[20]{MHz} sampling rate. About 100 measurement runs ---with 20,000 events each--- were performed with T2K gas (95\% Ar - 3\% CF$_4$ - 2\% isobutane) and with and without magnetic field of \unit[1]{T}.

\subsection{Module high voltage layout}

Two issues concerning the high voltage design of the module were identified in this module iteration. 

\begin{wrapfigure}{r}{0.40\columnwidth}
 \centerline{\includegraphics[width=0.32\columnwidth]{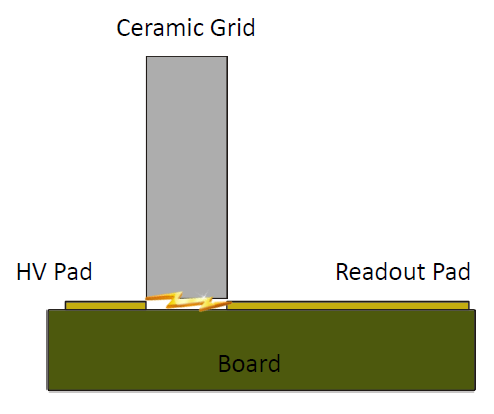}}
\caption [HVPads]{Schematic of the ceramic grid separating the high voltage and read-out pads.}
\label{fig:hvpads}
\end{wrapfigure}

The first issue was identified in tests before the test beam campaign and resulted from an insufficient insulation between the pads on the edge of the module, which feed the high voltage for the GEMs through the pad board, and the read-out pads. These are separated by the ceramic grid of the GEM mounting (Figure \ref{fig:hvpads}). Although the distance between the pads should have been sufficient to stand the applied voltages, sparks occurred during first high voltage tests. This could be solved by gluing the ceramic grid to the pad plane which sufficiently increased the insulation.

The second issue was only identified in the test beam setup. During the test beam campaign, the module showed a normal, rather low rate of GEM discharges. However, several of these discharges were powerful enough to destroy several GEM sectors. The most probable cause for this effect is the location of the protection resistors in the cables that supply the high voltage of the GEMs. In the setup where the module was tested before the campaign, these resistors were close to the GEMs, while in the test beam setup they were located farther away. The sum of the additional charge stored in the coaxial cables and the charge stored in the GEM sectors was large enough to burn short circuits in the GEM holes in case of a discharge.

\subsection{First data analysis}

In the following, a first, simple analysis of the data taken during the test beam campaign is presented. For this, eleven measurement runs taken without magnetic field at different drift lengths have been selected. 

\begin{figure}[hbt!]
\centering
\includegraphics[height=0.60\textwidth]{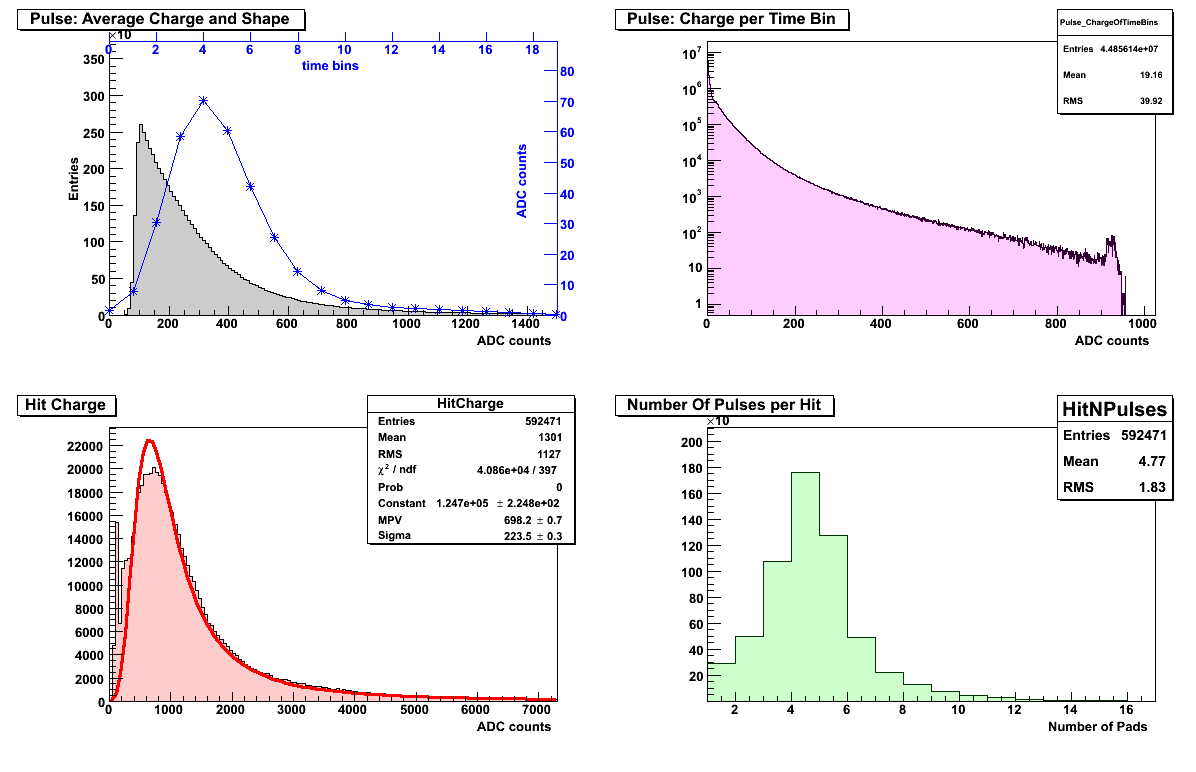}
\caption [SummaryPlots]{Top left: pulse charge and shape. Top right: charge per pulse time bin. Bottom left: charge per hit. Bottom right: number of pulses per hit.}
\label{fig:run17697}
\end{figure}

Figure \ref{fig:run17697} shows exemplarily for one run ---taken at a drift length of about \unit[185]{mm}--- four plots illustrating the quality of the recorded data. The plot on the top left shows in gray color a histogram of the pulse charge in ADC counts averaged over all pads. Here, {\em pulse} denotes the recorded charge signal on one pad. In blue the average shape of a pulse is shown in units of ADC counts versus the time in bins of \unit[50]{ns}. Both shapes correspond to the expectation. The top right plot shows a histogram of the charge of the single time bins of a pulse on a logarithmic scale. Due to the automatic pedestal subtraction of the read-out electronics, the highest values are below the maximum value of 1024. As expected, most pulse bins have a charge well below this maximum. 
The bottom left shows a histogram of the charge collected in a pad row. The combined signal of pulses in one row belonging to the same charge deposition is called a {\em hit}. The distribution follows the expected Landau like shape. The bottom right plot shows a histogram with the number of pulses per hit, showing that there is sufficient charge sharing with most hits spanning over 4 or more pads.

\begin{figure}[hbt!]
\centering
\includegraphics[height=0.26\textwidth,width=0.43\textwidth]{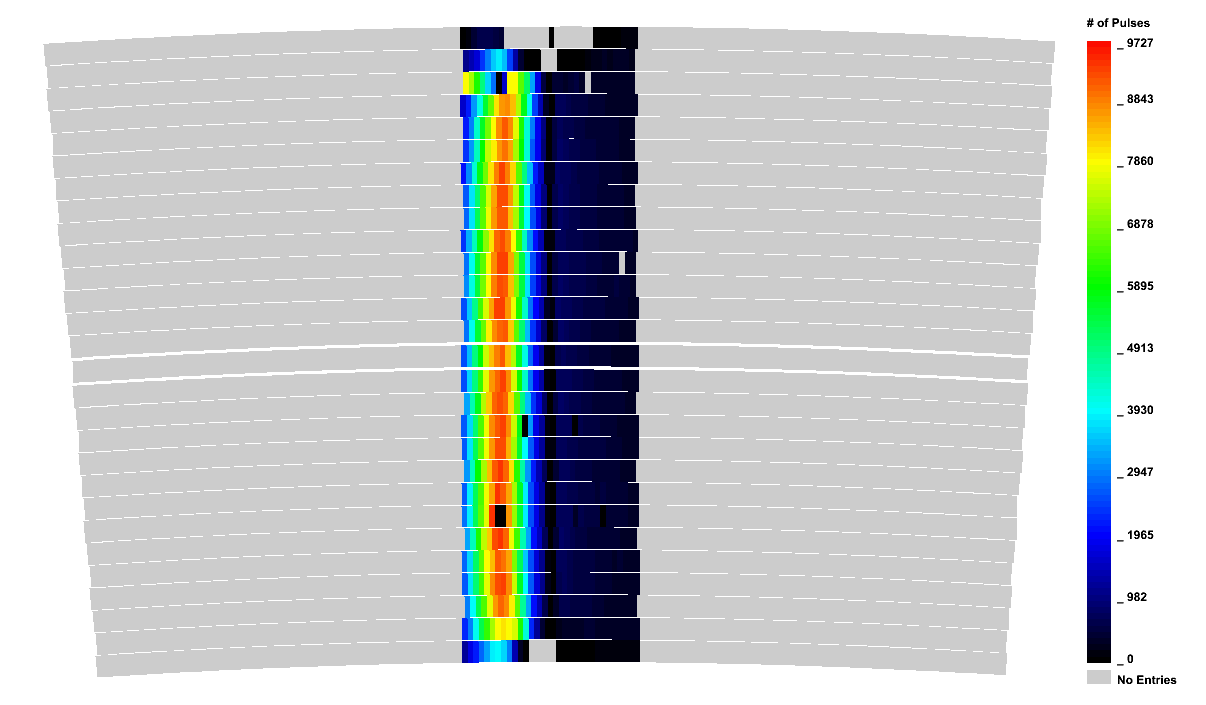}
\hspace{0.02\textwidth}
\includegraphics[height=0.26\textwidth,width=0.40\textwidth]{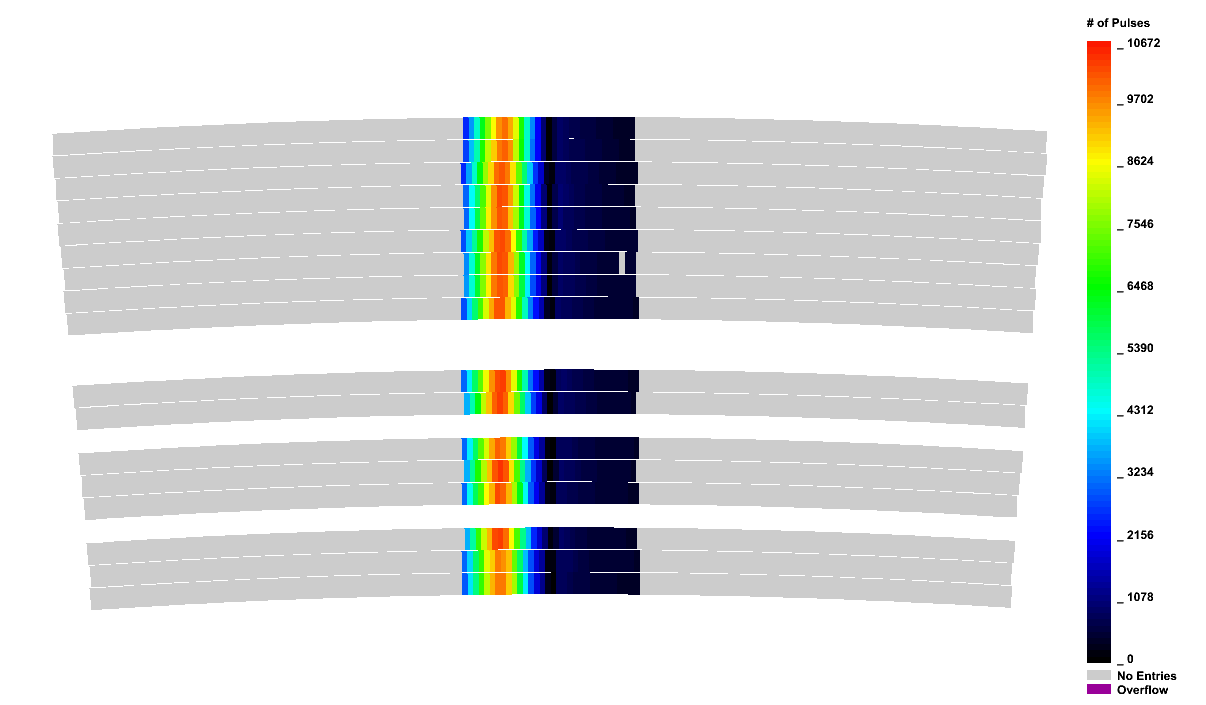}
\caption [PadPlaneSignal]{Sum of reconstructed pulses per read-out pad. Left: all pads before cut on data. Right: pads used in the analysis.}
\label{fig:padplane}
\end{figure}

In the following analysis, a cut on the pad row of the signal has been applied to ensure a good data quality. The rows 1 and 2 as well as the rows 24 to 28 (counted from the bottom) are not taken into account. In these rows, only a few signals were recorded due to field distortions at the edges of the module. In addition, signals from the rows 6,10 (contain dead channels) and 13,14 (close to the ceramic grid) were cut. Figure \ref{fig:padplane} illustrates these cuts.

\begin{figure}[hbt!]
\centering
\includegraphics[height=0.32\textwidth]{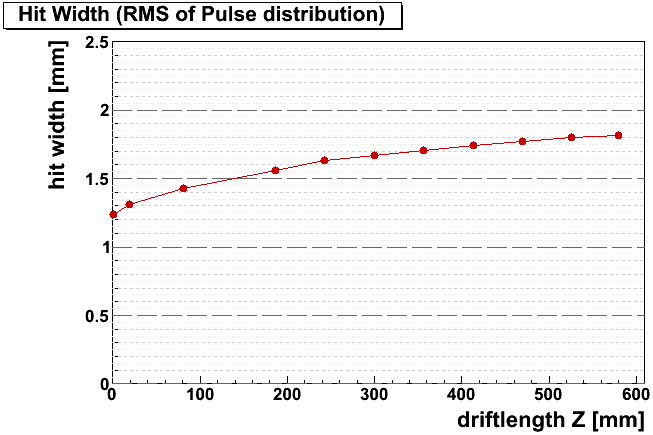}
\hspace{0.01\textwidth}
\includegraphics[height=0.32\textwidth]{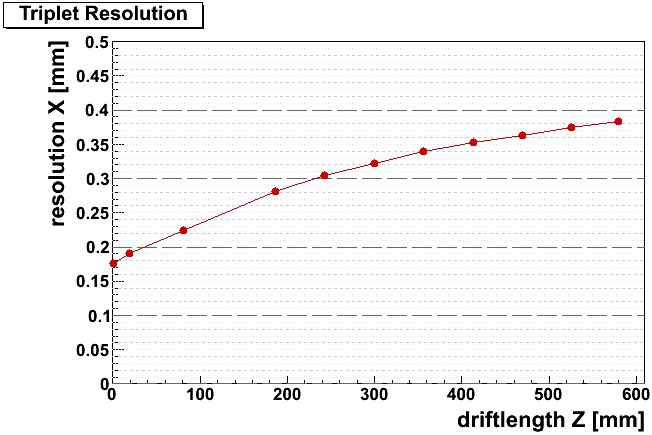}
\caption [HitWidthandResolution]{Left side: Hit width (RMS of pulse distribution). Right side: Triplet point resolution.}
\label{fig:hitwidthandresolution}
\end{figure}

In a further analysis step, the dependence of the hit width on the drift length and a first estimation of the point resolution has been investigated. The results are shown in Figure \ref{fig:hitwidthandresolution}. 

The RMS of the distribution of the pulse charges of a hit has been taken as a measure for its width. The mean values of the resulting RMS distribution of the measurement runs are plotted in Figure \ref{fig:hitwidthandresolution} against the drift length. The points in the plot follow the expected square root function shape. 

The point resolution has been determined applying the so-called {\em triplet method}. Here, the residual of a hit is defined as the distance of the hit from a straight line that connects the hits in the row above and below. The resolution is calculated from the width of the Gaussian residual distribution: $\sigma_{\mathrm{resolution}} = \sqrt{\frac{2}{3}} \sigma_{\mathrm{residual}}$. This allows for an estimate of the single point resolution independent of a track reconstruction. The resulting values match the expectation for data taken without magnetic field.

\section{Conclusion and Outlook}

A TPC read-out module using a novel GEM mounting technique using ceramic grids has been developed and tested in a test beam campaign. Several measurement runs with and without magnetic field were performed and a first analysis shows reasonable results. A more detailed analysis of the data is ongoing.

This test helped to identify the shortcomings of the current module design. Based on this experience, a new iteration of the module is being developed. The new iteration will include a read-out board of which the whole area is equipped with sensitive pads. The high voltage distribution is being redesigned to ensure a higher robustness and a better protection of the GEMs in case of discharges. Further, the use of plug connectors and the concentration of high voltage channels will improve the usability and operational reliability. In addition, techniques to minimize the field distortions at the module edges are investigated.

Last but not least, the test beam campaign and the following analysis triggered several improvements of the software framework including an improved event display, quick analysis programs and several bug fixes, improvements and additions to the reconstruction chain.
 
\section{Acknowledgments}

We would like to thank the LCTPC collaboration members from Lund University and from Japan for their invaluable help and support during the test beam campaign.

This work was supported in part by the U.S. National Science Foundation, by the  Commission of the European Communities under the 6th Framework Program ``Structuring the European Research Area" under contract RII3-026126 and the 7th Framework Programme ``Marie Curie ITN'', grant agreement number 214560, by the Swedish Council (VR) and by the HGF Alliance ``Physics at the Terascale''.

\section{Bibliography}

\begin{footnotesize}

\end{footnotesize}


\begin{thebibliography}{99}

\bibitem{:2010zzd}
  T.~Abe {\it et al.}  [ILD Concept Group - Linear Collider Collaboration],
  arXiv:1006.3396 [hep-ex].

\bibitem{ref-tpcprc2010} LCTPC Collaboration, {\it The Linear Collider Time Projection Chamber}, Report to the DESY PRC 2010,\\ http://www.lctpc.org/e10/e96773/e101340/infoboxContent102627/prc-report2010\_v19.pdf.

\bibitem{ref-lctpc} {{Homepage of the LCTPC collaboration, http://www.lctpc.org}}.

\bibitem{ref-eudet} {{Homepage of the EUDET collaboration, http://www.eudet.org/}}.

\bibitem{Behnke:2010ze}
  T.~Behnke, K.~Dehmelt, R.~Diener, L.~Hallermann, T.~Matsuda, V.~Prahl and P.~Schade,
  JINST {\bf 5}, P10011 (2010)
  [arXiv:1006.3220 [physics.ins-det]].

\bibitem{ref-endplate} {{D.~Peterson, web page for LP TPC end plate, http://www.lepp.cornell.edu/\symbol{126}dpp/linear\_collider/}}.

\bibitem{Hallermann:2010zz}
  L.~Hallermann,
  DESY-THESIS-2010-015.

\bibitem{ref-terascale} {{Homepage of the Helmholtz Alliance ``Physics at the Terascale'', http://www.terascale.de}}.

\bibitem{altro} L.~Musa et al., {\it The ALICE TPC Front End Electronics}, Proc. of the IEEE Nuclear Science
Symposium, 20-25.Oct.2003, Portland

\bibitem{pca16} L.~Musa, {\it Prototype compact readout system}, EUDET-Memo-2009-31, 2009

\end{thebibliography}
\end{document}